\documentclass{article}%
\usepackage{amsmath}
\usepackage{amsfonts}
\usepackage{amssymb}
\usepackage{graphicx}%
\newcommand{\bpartial}{\mathop{\partial\kern -4pt\raisebox{.8pt}{$|$}}}
\newcommand{\bra}{\mathopen{[\kern-1.6pt[}}
\newcommand{\ket}{\mathclose{]\kern-1.5pt]}}
\newcommand{\bbra}{\mathopen{[\kern-2.2pt[\kern-2.3pt[}}
\newcommand{\bket}{\mathclose{]\kern-2.1pt]\kern-2.3pt]}}

\makeindex
\begin{document}

\title {\large{ \bf (1+1)-dimensional gauge symmetric gravity model and related exact black hole and cosmological solutions in string theory }}

\vspace{3mm}

\author {  \small{ \bf S. Hoseinzadeh }\hspace{-2mm}{ \footnote{ e-mail: hoseinzadeh@azaruniv.edu}} { \small
and}
\small{ \bf A. Rezaei-Aghdam }\hspace{-2mm}{ \footnote{Corresponding author. e-mail:
rezaei-a@azaruniv.edu}} \\
{\small{\em Department of Physics, Faculty of Science, Azarbaijan Shahid Madani University, }}\\
{\small{\em  53714-161, Tabriz, Iran  }}}

\maketitle

\begin{abstract}
We introduce a four-dimensional extension of the Poincar\'{e}
algebra $(\mathcal{N})$ in (1+1)-dimensional space-time and
obtain a (1+1)-dimensional gauge symmetric gravity model using the algebra
$\mathcal{N}$. We show that the obtained gravity model is dual
(canonically transformed) to the (1+1)-dimensional anti de Sitter ($AdS$) gravity. We also obtain some black hole and Friedmann-Robertson-Walker (FRW) solutions by solving its classical equations of motion. 
Then, we study $\frac{\mathbf{A}_{\mathbf{4,8}}}{\mathbf{A}_{\mathbf{1}}\otimes \mathbf{A}_{\mathbf{1}}}$ gauged Wess-Zumino-Witten (WZW) model and obtain some exact black hole and cosmological solutions in string theory. We show that some obtained black hole and cosmological metrics in string theory are same as the metrics obtained in solutions of our gauge symmetric gravity model.

\end{abstract}
\newpage
%%%%%%%%%%%%%%%%%%%%%%%%%%%%%%%%%%%%%%%%%%%%%%%%%%%%%%%%%%%%%%%%%%%%%%%%%%%%%%%%%%%%%%%%%%%%%%%%%%%%%%%%%%
\section {\large {\bf Introduction}}
\setlength{\parindent}{0cm}

(1+1)-dimensional gravity has been extensively studied by proposing various models. 
Two of the gravitational theories of most interest are singled out by their simplicity and group theoretical properties. One of them is proposed by Jackiw \cite{1Jackiw} and Teitelboim \cite{Teitelboim} (Liouville gravity) which is equivalent to the gauge theory of gravity with (anti) de Sitter group \cite{Fukuyama,Isler,Chamseddine}. The other one is the string-inspired gravity \cite{Witten,Verlinde,1Callan} which is equivalent to the gauge theory of the Poincar\'{e} group $ISO(1,1)$  \cite{Verlinde} and its central extension 
\cite{1Cangemi,Jackiw,2Cangemi,Achucarro,Grignani}.

Recently, two algebras namely the Maxwell algebra
\cite{Bacry,Schrader} and the semi-simple extension of the Poincar\'{e} algebra \cite{Soroka} have been applied to construct some gauge invariant theories of gravity in four \cite{Soroka,1Azcarraga,2Azcarraga,1Cebecioglu,2Cebecioglu} and three \cite{Salgado,Diaz,1Hoseinzadeh} dimensional space-times. These algebras have been also applied to string theory as an internal symmetry of the matter gauge fields
\cite{2Hoseinzadeh}. The Maxwell algebra in (1+1)-dimensional space-time, is the well-known central extension of the Poincar\'{e} algebra which, as we discussed above, has been
applied to construct a (1+1)-dimensional gauge symmetric gravity action \cite{1Cangemi,Jackiw}. In this paper, we introduce a new four-dimensional extension of the Poincar\'{e} algebra $(\mathcal{N})$ in (1+1)-dimensional space-time, which is obtained from the $16$-dimensional semi-simple extension of Poincar\'{e} algebra in (3+1)-dimensional space-time \cite{Soroka}, by reduction of the dimensions of the space. Then, we construct a (1+1)-dimensional gauge symmetric gravity model, using this algebra. We obtain some black hole and cosmological solutions by solving its equations of motion.

On the other hand, in string theory, two-dimensional exact black hole has been found by Witten \cite{Witten}. Another black hole solution to the string theory has been presented in \cite{Mandal} both in Schwarzschild-like and target space conformal gauges. Exact three-dimensional black string and black hole solutions in string theory have also been found in \cite{Horne,Horowitz}. Here, we study the string theory in (1+1)-dimensional space-time, and show that some obtained black hole and cosmological solutions of the gravity model, are exact solutions of the beta function equations (in all loops).

The outline of this paper is as follows: In section two, we construct a (1+1)-dimensional gauge symmetric gravity model using a four-dimensional gauge group related to the algebra
$\mathcal{N}$. Then, by presenting a canonical map, we show that the obtained gravity model is dual (canonically transformed) to the (1+1)-dimensional $AdS$ gravity model. In section three, we solve the equations of motion and obtain some black hole and Friedmann-Robertson-Walker (FRW) cosmological solutions. Finally, in section four, we study $\frac{\mathbf{A}_{\mathbf{4,8}}}{\mathbf{A}_{\mathbf{1}}\otimes \mathbf{A}_{\mathbf{1}}}$ gauged Wess-Zumino-Witten (WZW)
model, and show that some of the resulting string backgrounds, which are exact (1+1)-dimensional solutions of the string theory, are the same as the black hole and cosmological solutions obtained for our gravity model. Section five, contains some concluding remarks.

%%%%%%%%%%%%%%%%%%%%%%%%%%%%%%%%%%%%%%%%%%%%%%%%%%%%%%%%%%%%%%%%%%%%%%%%%%%%%%%%%%%%%%%%%%%%%%%%%%%%%%%%%%
\section {\large {\bf (1+1)-dimensional gravity from a non-semi-simple extension of the Poincar\'{e} gauge symmetric model}}
\setlength{\parindent}{0cm}

The Poincar\'{e} algebra $I\!so(1,1)$ in (1+1)-dimensional
space-time has the following form:
\begin{equation}
[J,P_{a}]=\epsilon_{ab}P^{b},~~~~~~~~~ [P_{a},P_{b}]=0,
\end{equation}
where $\epsilon_{01}=-\epsilon^{01}=+1,$ and  $J$  and  $P_{a}$
$(a=0,1)$ are generators of the rotation and translations in
space-time, and the algebra indices $a=0,1$ can be raised and lowered by the (1+1)-dimensional Minkowski metric $\eta_{ab}$ ($\eta_{00}=-1, \eta_{11}=1$) such that $P_{a}=\eta_{ab}P^{b}$.
In $D=1+1$, a four-dimensional non-semi-simple
extension of the Poincar\'{e} algebra\footnote{This algebra is
isomorphic to the four-dimensional Lie algebra $\mathcal{A}_{3,8} \oplus
\mathcal{A}_{1}$ \cite{Patera}.} $\mathcal{N}=(P_{a},J,Z)$ has the following form:
\begin{equation}
[J,P_{a}] = \epsilon_{ab}P^{b},~~~~~~ [P_{a},P_{b}] = k
\epsilon_{ab}Z,~~~~~~
[Z,P_{a}]=-\frac{\Lambda}{k}\epsilon_{ab}P^{b},
\end{equation}
where $Z$ is the new generator and $k$ and $\Lambda$ are
constants.\footnote{Note that the commutation relation $[J,Z]=0$ can be
obtained from the Jacobi identity $[J,[P_{a},P_{b}]]+$ cyclic terms $=0$.} For $\Lambda=0,$ which
leads to $[Z,P_{a}]=0$, the above algebra reduces to a solvable
algebra which is called the centrally extended Poincar\'{e} algebra (or Maxwell algebra\footnote{Centrally extended Poincar\'{e} algebra (or Maxwell algebra) in 1+1 dimensions is isomorphic to the four-dimensional Lie algebra $\mathcal{A}_{4,8}$ \cite{Patera}.} in 1+1 dimensions)
\cite{1Cangemi,Jackiw,2Cangemi}. We construct the
$\mathcal{N}$-algebra valued one-form gauge field as follows:
\begin{equation}
h_{i}=h_{i}^{~B}X_{B}=e_{i}^{~a}P_{a}+\omega_{i}~J+A_{i}~Z, ~~~~~~~~ i,j=0,1
\end{equation}
where the indices $i,j=0,1$ are the space-time indices, and the
one-form fields have the following forms:
\begin{equation} \nonumber
e^{a}=e_{i}^{~a} dx^{i} ~~,~~ \omega=\omega_{i} dx^{i} ~~,~~ A=A_{i} dx^{i},
\end{equation}
where $ e_{i}^{~a}, \omega_{i}, A_{i} $ are the vierbein, spin
connection and the new gauge field, respectively.
Using the following infinitesimal gauge parameter:
\begin{equation} \nonumber
u=\rho^{a}P_{a}+\tau~J+\lambda~Z,
\end{equation}
and the gauge transformation as follows:
\begin{equation} \nonumber
h_{i} \rightarrow h^{\prime}_{i}=U^{-1}h_{i}U+U^{-1}\partial_{i}U ,
\end{equation}
with ~$U = e^{-u}~ \simeq ~ 1-u$ ~and~ $U^{-1} = e^{u}~ \simeq ~ 1+u,$
we obtain the following transformations of the gauge fields:
\begin{equation}    \nonumber
\delta e_{i}^{\ a}=-\partial_{i}\rho^{a}-\epsilon^{ab} e_{ib}~
(\tau -\frac{\Lambda}{k}~\lambda) +\epsilon^{ab}
\rho_{b}~(\omega_{i} -\frac{\Lambda}{k}A_{i})~,
\end{equation}
\begin{equation}\label{Transformations of gauge fields}
\delta \omega_{i}=-\partial_{i}\tau, ~~~~~~~~~~~~~~~~~~~~~~~~~~~~~~~~~~~~~~~~~~~~~~~~~~~~~
\end{equation}
\begin{equation}     \nonumber
\delta A_{i}=-\partial_{i}\lambda -k~\epsilon^{ab} e_{ia} ~ \rho_{b}.~~~~~~~~~~~~~~~~~~~~~~~~~~~~~~~~~~~~
\end{equation}
The generic Ricci curvature can be obtained as follows:
\begin{equation}\nonumber
{\cal R}={\cal R}_{ij} dx^{i} \wedge dx^{j} = {\cal R}^{A} X_{A}
={\cal R}_{ij}^{~A} X_{A} dx^{i} \wedge dx^{j} ,~~~~~~~~~~~~
\end{equation}
\begin{equation}
{\cal R}_{ij}=\partial_{[i}h_{j]}+[h_{i},h_{j}] = {\cal R}_{ij}^{~A} X_{A} =T_{ij}^{~~a}~P_{a}+R_{ij}~J+F_{ij}~Z ~,
\end{equation}
where the torsion $T_{ij}^{~~a}$, standard Riemannian curvature
$R_{ij}$ and the new gauge field strength $F_{ij}$ have the
following forms:
\begin{equation}  \nonumber
T_{ij}^{~~a}=\partial_{[i}~e_{j]}^{~a}+\epsilon^{ab}(e_{ib}~\omega_{j}-
e_{jb}~\omega_{i}) -\frac{\Lambda}{k}~\epsilon^{ab}(e_{ib}~A_{j}-
e_{jb}~A_{i}),
\end{equation}
\begin{equation}\label{Fields strength}
R_{ij}=\partial_{[i}~\omega_{j]},~~~~~~~~~~~~~~~~~~~~~~~~~~~~~~~~~~~~~~~~~~~~~~~~~~~~~~~~~~~~
\end{equation}
\begin{equation}  \nonumber
F_{ij}=\partial_{[i}~A_{j]} + k~\epsilon^{ab} e_{ia}~e_{jb}.~~~~~~~~~~~~~~~~~~~~~~~~~~~~~~~~~~~~~~~~~
\end{equation}
Now, one can write the gauge invariant action as \cite{Jackiw}
\begin{equation}
I= \frac{1}{2} \int \eta _{A}{\cal R}^{A}= \frac{1}{2} \int d^{2}x ~\epsilon^{ij}~\eta_{A}~{\cal R}^{~~A}_{ij}
\end{equation}
\begin{equation}
~~~~~~~~= \frac{1}{2} \int d^{2}x~\epsilon^{ij}~ (\eta_{a}~T_{ij}^{~~a} +\eta_{2}~R_{ij} +\eta_{3}~F_{ij}),
\end{equation}
where $\eta_{A}=(\eta_{a},\eta_{2},\eta_{3})$ are the lagrange
multiplier-like fields. Now, using \eqref{Fields strength}, one can rewrite this action in the following form:
\begin{equation}\label{Main action}
I= \int d^{2}x ~  \epsilon^{ij}~ \Big\{ \eta_{a}~
\Big(\partial_{i} e_{j}^{~a} + \epsilon^{ab}~ e_{ib}~ (\omega_{j}
-\frac{\Lambda}{k}~ A_{j})\Big) + \eta_{2}~\partial_{i}
\omega_{j} + \eta_{3}~\Big(\partial_{i} A_{j}
+\frac{1}{2}k~\epsilon^{ab}~ e_{ia}~ e_{jb}\Big) \Big\}.
\end{equation}
This action is invariant under the gauge transformations
\eqref{Transformations of gauge fields} and the following transformations of the fields
$\eta_{a}$, $\eta_{2}$ and $\eta_{3}$:
\begin{equation}\nonumber
\delta \eta_{a}=k~\epsilon_{a}^{~~b}~\eta_{3}~\rho_{b}
-\epsilon_{a}^{~~b}~\eta_{b} (\tau -\frac{\Lambda}{k}~\lambda),
\end{equation}
\begin{equation}
\delta \eta_{2}= -\epsilon^{ab} \eta_{a}~\rho_{b},~~~~~~~~~~~~~~~~~~~~~~~~~
\end{equation}
\begin{equation}\nonumber
\delta \eta_{3}= \frac{\Lambda}{k}~ \epsilon^{ab} \eta_{a}~\rho_{b}.~~~~~~~~~~~~~~~~~~~~~~~~
\end{equation}
Now, we will show that the model \eqref{Main action} is dual to the (1+1)-dimensional $AdS$ gravity.
We know that $SO(2,1)$ gauge symmetric gravity action can be obtained by use of the following algebra (anti de Sitter algebra for $k'\neq0$):
\begin{equation}
[J,P_{a}]=\epsilon_{ab}P^{b},~~~~~~~~~
[P_{a},P_{b}]=k'\epsilon_{ab}J,
\end{equation}
as follows: \cite{Jackiw}
\begin{equation}\label{AdS gravity}
\tilde{I}= \int d^{2}x ~  \epsilon^{ij}~ \Big\{
\widetilde{\eta}_{a}~ \Big(\partial_{i} e_{j}^{~a} +
\epsilon^{ab}~ e_{ib}~ \omega_{j} \Big)
+\widetilde{\eta}_{2}~\Big(\partial_{i} \omega_{j}
+\frac{1}{2}k'~\epsilon^{ab}~ e_{ia}~ e_{jb}\Big) \Big\}.
\end{equation}
An $SO(2,1)$ invariant action for two-dimensional gravity was first constructed in \cite{Fukuyama} where the aim was to reconstruct the proposed two-dimensional Einstein equation from a two-dimensional gauge theory of gravity. Although the notation adopted in \cite{Fukuyama} is different from our notation,\footnote{The $SO(2,1)$ invariant action for two-dimensional gravity constructed in \cite{Fukuyama} is $\frac{1}{2}\!\int \!\epsilon^{A\!BC} R_{A\!B}~\phi_{C}$ where $R_{A\!B}=d\omega_{A\!B}-\omega_{AC}~\omega^{C}_{~B}$ and $A,B=0,1,2$. The field $\omega_{AB}$ contains both the spin connection $\omega_{ab}$ and vierbein $e_{a}$ where $a,b=0,1$ such that we have $\omega_{a2}=\ell^{-1} e_{a}$.  Using the field redefinitions  $\phi_{0}\equiv\ell\tilde{\eta}_{1}$, $\phi_{1}\equiv\ell\tilde{\eta}_{0}$, $\phi_{2}\equiv\tilde{\eta}_{2}$ and $\omega_{01}\equiv\omega$, one can easily show that this action is equivalent to the $AdS_{2}$ gravity action \eqref{AdS gravity} with $k'=\ell^{-2}$.} but it can be shown that the action constructed in \cite{Fukuyama} is equivalent to the (1+1)-dimensional $AdS$ gravity model \eqref{AdS gravity}.
Now, by selecting $\eta_{3}=-\frac{\Lambda}{k}\eta_{2}$ in our model \eqref{Main action}, it is dual (canonically transformed) to the $AdS$ gravity \eqref{AdS gravity}; i.e. the
following map:
\begin{equation}\label{Map}
\omega_{i}\longrightarrow \omega_{i}- \frac{\Lambda}{k}
A_{i},~~~~~~~ e_{i}^{~a}\longrightarrow e_{i}^{~a},~~~~~~~
\widetilde{\eta}_{a}\longrightarrow
\eta_{a},~~~~~~~\widetilde{\eta}_{2}\longrightarrow
\eta_{2},~~~~~~~ k'=-\Lambda,
\end{equation}
transforms the $AdS_{2}$ gravity model \eqref{AdS gravity} to our model \eqref{Main action}. In the following, we will show that this map is a canonical one. The canonical Poisson-brackets and the
Hamiltonian related to the $AdS_{2}$ gravity model \eqref{AdS gravity} are
as follows:
\begin{equation}\nonumber
\{(\tilde{\Pi}_{e})_{i}^{~a}(x) ~,~ e_{j}^{~b}(y)\}=\epsilon_{ij} \eta^{ab} \delta(x-y),~~~~~~~~~
\{(\tilde{\Pi}_{\omega})_{i}(x) ~,~ \omega_{j}(y)\}=
\epsilon_{ij} \delta(x-y),
\end{equation}
\begin{equation}\nonumber
\{(\tilde{\Pi}_{\widetilde{\eta}_{a}})^{a}(x) ~,~
\widetilde{\eta}^{b}(y)\}= \eta^{ab} \delta(x-y),~~~~~~~~~
\{(\tilde{\Pi}_{\widetilde{\eta}_{2}})(x) ~,~
\widetilde{\eta}_{2}(y)\}= \delta(x-y),
\end{equation}
\begin{equation}\label{Hamiltonian}
\tilde{H}= \int d^{3}x\Big( (\tilde{\Pi}_{e})^{i}_{~a}~ \partial_{t}e_{i}^{~a} + (\tilde{\Pi}_{\omega})^{i}_{~a}~ \partial_{t}\omega_{i}^{~a}+(\tilde{\Pi}_{\widetilde{\eta}_{a}})^{a}~ \partial_{t}\widetilde{\eta_{a}} + (\tilde{\Pi}_{\widetilde{\eta}_{2}})~ \partial_{t}\widetilde{\eta}_{2} \Big) - \tilde{I}
\end{equation}
\begin{equation}\nonumber
= 2\int d^{2}x \epsilon^{0i} \Big(
\widetilde{\eta}_{a}
\partial_{t} e_{i}^{~a} + \widetilde{\eta}_{2}
\partial_{t} \omega_{i}  \Big) - \tilde{I},
\end{equation}
where the coordinates of the space-time are $\{x^{i}\}=\{x^{0},x^{1}\}=\{t,r\}$ such that $\partial_{t}=\partial_{0}=\frac{\partial}{\partial t}$, $\eta^{ab}$ is the inverse Minkowski metric, and the conjugate momentums corresponding to the fields are as follows:
\begin{equation}\nonumber
(\tilde{\Pi}_{e})^{i}_{~a}=
\frac{\partial \mathcal{\tilde{L}}}{\partial (\partial_{t}e^{~a}_{i})}= \epsilon^{0i} \widetilde{\eta}_{a},~~~~~~~~ (\tilde{\Pi}_{\omega})^{i}=
\frac{\partial \mathcal{\tilde{L}}}{\partial (\partial_{t}\omega_{i})}= \epsilon^{0i} \widetilde{\eta}_{2},
\end{equation}
\begin{equation}
(\tilde{\Pi}_{\widetilde{\eta}_{a}})^{a}=
\frac{\partial \mathcal{\tilde{L}}}{\partial (\partial_{t}\widetilde{\eta}_{a})}= -\epsilon^{0i} e_{i}^{~a},~~~~~~~~ (\tilde{\Pi}_{\widetilde{\eta}_{2}})=
\frac{\partial \mathcal{\tilde{L}}}{\partial (\partial_{t}\widetilde{\eta}_{2})}= -\epsilon^{0i} \omega_{i}.
\end{equation}
The map \eqref{Map} is a canonical transformation and easily it
can be shown that, under this map, the canonical Poisson-brackets
and the Hamiltonian \eqref{Hamiltonian} related to the $AdS_{2}$ gravity
model \eqref{AdS gravity} are transformed to the following
Poisson-brackets and the Hamiltonian related to our model
\eqref{Main action}:
\begin{equation}\nonumber
\{(\Pi_{e})_{i}^{~a}(x) ~,~ e_{j}^{~b}(y)\}=\epsilon_{ij}
\eta^{ab} \delta(x-y),~~~~~~~~~ \{(\Pi_{\omega})_{i}(x) ~,~
\omega_{j}(y)\}= \epsilon_{ij} \delta(x-y),
\end{equation}
\begin{equation}\nonumber
\{(\Pi_{A})_{i}(x) ~,~ A_{j}(y)\}= \epsilon_{ij}
\delta(x-y),~~~~~~~~~ \{(\Pi_{\eta_{a}})^{a}(x) ~,~
\eta^{b}(y)\}= \eta^{ab} \delta(x-y),
\end{equation}
\begin{equation}\nonumber
\{(\Pi_{\eta_{2}})(x) ~,~ \eta_{2}(y)\}= \delta(x-y),~~~~~~~~~
\{(\Pi_{\eta_{3}})(x) ~,~ \eta_{3}(y)\}= \delta(x-y),
\end{equation}
\begin{equation}
H= \int d^{3}x\Big( (\Pi_{e})^{i}_{~a}~
\partial_{t}e_{i}^{~a} + (\Pi_{\omega})^{i}_{~a}~
\partial_{t}\omega_{i}^{~a}+ (\Pi_{A})^{i}_{~a}~
\partial_{t}A_{i}^{~a}+(\Pi_{\eta_{a}})^{a}~
\partial_{t}\eta_{a} +(\Pi_{\eta_{2}})~
\partial_{t}\eta_{2}+(\Pi_{\eta_{3}})~
\partial_{t}\eta_{3} \Big) -I
\end{equation}
\begin{equation}\nonumber
= 2\int d^{2}x \epsilon^{0i} \Big( \eta_{a}
\partial_{t} e_{i}^{~a} +\eta_{2}
\partial_{t} \omega_{i} +\eta_{3}
\partial_{t} A_{i} \Big) -I,
\end{equation}
where the conjugate momentums corresponding to the fields are
given as follows:
\begin{equation}\nonumber
(\Pi_{e})^{i}_{~a}= \frac{\partial \mathcal{L}}{\partial
(\partial_{t}e^{~a}_{i})}= \epsilon^{0i}
\eta_{a},~~~~~~~~ (\Pi_{\omega})^{i}= \frac{\partial
\mathcal{L}}{\partial (\partial_{t}\omega_{i})}= \epsilon^{0i}
\eta_{2},
\end{equation}
\begin{equation}
(\Pi_{A})^{i}= \frac{\partial \mathcal{L}}{\partial
(\partial_{t}A_{i})}= \epsilon^{0i} \eta_{3},~~~~~~~~
(\Pi_{\eta_{a}})^{a}= \frac{\partial \mathcal{L}}{\partial
(\partial_{t}\eta_{a})}= -\epsilon^{0i} e_{i}^{~a},
\end{equation}
\begin{equation} \nonumber
(\Pi_{\eta_{2}})= \frac{\partial \mathcal{L}}{\partial
(\partial_{t}\eta_{2})}= -\epsilon^{0i} \omega_{i},~~~~~~~~
(\Pi_{\eta_{3}})= \frac{\partial \mathcal{L}}{\partial
(\partial_{t}\eta_{3})}= -\epsilon^{0i} A_{i}.
\end{equation}
Note that the conjugate momentums are transformed under the map
\eqref{Map} as:
\begin{equation}\nonumber
(\widetilde{\Pi}_{e})^{i}_{~a} \longrightarrow
(\Pi_{e})^{i}_{~a},~~~~~~~
(\widetilde{\Pi}_{\widetilde{\eta}_{a}})^{a} \longrightarrow
(\Pi_{\eta_{a}})^{a},
\end{equation}
\begin{equation} \nonumber
(\widetilde{\Pi}_{\omega})^{i} \longrightarrow
(\Pi_{\omega})^{i},~~~~~~~ \widetilde{\Pi}_{\widetilde{\eta}_{2}}
\longrightarrow (\Pi_{\eta_{2}}-\frac{\Lambda}{k}\Pi_{\eta_{3}}).
\end{equation}
Since the Poisson-brackets and Hamiltonian of the model are
preserved under the map \eqref{Map}, then the $AdS$ gauge gravity
model \eqref{AdS gravity} is dual to our gravity model \eqref{Main action}, and
each can be transformed to the other by the canonical
transformation \eqref{Map}, of course with
$\eta_{3}=-\frac{\Lambda}{k}\eta_{2}$. Finally, under the map
\eqref{Map}, the equations of motion for the $AdS_{2}$ gravity
\eqref{AdS gravity} also transform to the equations of motion
related to our model \eqref{Main action}.

The equations of motion with respect to the fields $\eta_{a},
\eta_{2}, \eta_{3}$ have the following forms, respectively:
\begin{equation} \nonumber
\epsilon^{ij} \Big( \partial_{i} e_{j}^{~a} +\epsilon^{ab}~
e_{ib}~ (\omega_{j} -\frac{\Lambda}{k}~ A_{j}) \Big)=0,
\end{equation}
\begin{equation}\label{e.o.m.1}
\epsilon^{ij} \partial_{i}~\omega_{j}=0,
\end{equation}
\begin{equation} \nonumber
\epsilon^{ij} \Big( \partial_{i} A_{j}
+\frac{1}{2}k~\epsilon^{ab}~ e_{ia}~ e_{jb} \Big)=0,
\end{equation}
and the equations of motion with respect to the fields
$e_{i}^{~a}, \omega_{i}, A_{i}$ are obtained as follows, respectively:
\begin{equation} \nonumber
\epsilon^{ij} \Big( -\partial_{j} \eta_{a} +\epsilon_{a}^{~~b}~
\eta_{b}~ (\omega_{j} -\frac{\Lambda}{k}~ A_{j}) -
k~\epsilon_{ab}~ \eta_{3}~ e_{j}^{~b} \Big)=0,
\end{equation}
\begin{equation} \label{e.o.m.2}
\epsilon^{ij} \Big( \partial_{j} \eta_{2} -\epsilon^{ab}~
\eta_{a}~ e_{jb} \Big)=0,
\end{equation}
\begin{equation} \nonumber
\epsilon^{ij} \Big( \partial_{j} \eta_{3} +\frac{\Lambda}{k}
\epsilon^{ab}~ \eta_{a}~ e_{jb} \Big)=0.
\end{equation}
In the next section, we will try to solve these equations and obtain different solutions of them.

%%%%%%%%%%%%%%%%%%%%%%%%%%%%%%%%%%%%%%%%%%%%%%%%%%%%%%%%%%%%%%%%%%%%%%%%%%%%%%%%%%%%%%%%%
\section  {Solutions of the equations of motion}

\subsection {Radial Solutions for $\Lambda \neq 0$}
Using the following Ansatz for the metric:
\begin{equation}\label{Metric Ansatz}
ds^2=e_{i}^{~a}e_{j}^{~b}\eta_{ab}dx^{i}dx^{j}=- N^2(r)~dt^2+ M^2(r) dr^2,
\end{equation}
where $\{x^{0}, x^{1}\} = \{t, r\}$ are the coordinates of the
space-time $(0\leq t<\infty,~ -\infty<r<\infty)$, one can obtain the following
solution for the equations of motion \eqref{e.o.m.1}-\eqref{e.o.m.2}:
\begin{equation} \nonumber
M^2(r)=\frac{1}{-\Lambda N^{2}(r)+C_{4}} \Big(\frac{dN(r)}{dr}\Big)^{2},~~~~~~~~~~~~~  \eta_{0}(r)=C_{2}
N(r),~~~~~~~~~   \eta_{1}(r)=0,~~~~~~~~~~~~~~~~~~~~~~
\end{equation}
\begin{equation} \label{Solution for non-zero Lambda}
~~~~~~~~~~\eta_{2}(r)=\frac{C_{2}}{\Lambda} \sqrt{-\Lambda N^{2}(r)+C_{4}} +C_{1},~~~~~~~~~~~~~
\eta_{3}(r)=-\frac{C_{2}}{k} \sqrt{-\Lambda N^{2}(r)+C_{4}},~~~~~~~~~~~~~~~~~~~~~~~~~~~~~~~~~~
\end{equation}
\begin{equation} \nonumber
~~~~~~~~\omega(r)=C_{3} ~dt +f(r) ~dr,~~~~~~~~~~~~~~~~~~~~~~~~~~~~ 
A(r)=\frac{k}{\Lambda} \Big((C_{3}-\sqrt{-\Lambda N^{2}(r)+C_{4}}) ~dt +f(r) ~dr \Big),~~~~~~
\end{equation}
where $C_{1}$, $C_{2}$, $C_{3}$ and $C_{4}$ are arbitrary constants, and $N(r)$, $f(r)$
are arbitrary functions of ~$r$.
The solution \eqref{Solution for non-zero Lambda} describes a space-time with a constant scalar curvature $\mathcal{R} =2 \Lambda$.

\subsubsection {$AdS$ black hole solution}

For $N^{2}(r)=-\Lambda r^2 -b$ and $C_{4}=-\Lambda b$, the solution \eqref{Solution for non-zero Lambda} reduces to the following $AdS$ black hole solution:

\begin{equation} \label{Schwarzschild metric}
ds^2=- N^2(r)~dt^2+ \frac{dr^2}{N^2(r)},
\end{equation}
and
\begin{equation} \nonumber
\eta_{0}(r)=C_{2}
N(r),~~~~~~~~~~~~~~~~~~~~~~~~~~   \eta_{1}(r)=0,~~~~~~~~~~~~~~~~~~~~~~~~~~~~~~~~~~~~~~~~~~~
\end{equation}
\begin{equation} \label{AdS black hole solution}
\eta_{2}(r)=- C_{2} ~r +C_{1},~~~~~~~~~~~~~~~~~~~~~~~
\eta_{3}(r)=\frac{\Lambda}{k} C_{2}
~r,~~~~~~~~~~~~~~~~~~~~~~~~~~~~~~~~~~
\end{equation}
\begin{equation} \nonumber
\omega(r)=C_{3} ~dt +f(r) ~dr,~~~~~~~~~~~~~~~~~ A(r)=(k r
+\frac{k}{\Lambda}C_{3}) ~dt +\frac{k}{\Lambda}f(r) ~dr,~~~~~~
\end{equation}
where $b$ is a constant.
Now, we calculate the mass (energy) of solution \eqref{AdS black hole solution} using the ADM definition of mass (energy) as discussed in \cite{Bak}.
Varying the action \eqref{Main action} produces a bulk term, which is zero using the equations of motion, plus a boundary term which can be cancelled by adding the following boundary term to the Lagrangian:
\begin{equation}
{\cal L}_{B}=-~\partial_{r} \Big( \eta_{a} e_{0}^{~a} +\eta_{2} \omega_{0} +\eta_{3} A_{0} \Big),
\end{equation}
together with an appropriate boundary condition. This boundary term is identified as the mass (energy) of solution. Our boundary condition is using the obtained values for fields in the solution \eqref{AdS black hole solution} at spatial infinity ($r \rightarrow \pm \infty$). Then, the mass of the solution is obtained as follows:
\begin{equation}
m =\int_{-\infty}^{+\infty} dr {\cal L}_{B}=- \Big( \eta_{a} e_{0}^{~a} +\eta_{2} \omega_{0} +\eta_{3} A_{0} \Big) \Big{|}_{-\infty}^{+\infty},
\end{equation}
which using \eqref{Schwarzschild metric} and \eqref{AdS black hole solution} turns out to be
\begin{equation}
m =C_{2} b -C_{1}C_{3}.
\end{equation}
The Kretschmann scalar for this metric is
\begin{equation}
K = R_{\mu\nu\rho\sigma} R^{\mu\nu\rho\sigma}=4 \Lambda^{2}.
\end{equation}
Consequently, this solution has two singularities at the
following points:
\begin{equation}
r_{\pm}=\pm\sqrt{\frac{b}{-\Lambda}},
\end{equation}
which are not the curvature singularities, but the coordinate
singularities, and can be removed by definition of a new
coordinate system.
Using the Ansatz \eqref{Schwarzschild metric}, we obtain another solution for the
equations of motion \eqref{e.o.m.1}-\eqref{e.o.m.2} as follows:
\begin{equation} \nonumber
N^2(r)=- \Lambda r^2 -2D r +C_{3},~~~~~~~~~~~~  \eta_{0}(r)=C_{2}
N(r),~~~~~~~~~~   \eta_{1}(r)=0,~~~~~~~~~~~~~~~~~~~~
\end{equation}
\begin{equation}\label{AdS black hole solution-2}
\eta_{2}(r)=-C_{2} ~r +C_{1},~~~~~~~~~~~~~~~~~~~~~~~~~~~~~~~~~~~~
\eta_{3}(r)=\frac{C_{2}}{k}(\Lambda ~r +D),~~~~~~~~~~~~~~~~~~
\end{equation}
\begin{equation} \nonumber
\omega(r)=C_{4} ~dt +f(r) ~dr,~~~~~~~~~~~~~~~~~~~~~~~~~~~~~
A(r)=(k r +C_{5}) ~dt +\frac{k}{\Lambda}f(r)
~dr,~~~~~~~~~~~~~~~~~
\end{equation}
where $C_{1}$, $C_{2}$, $C_{3}$, $C_{4}$, $C_{5}$ and ~$D=
\frac{\Lambda}{k}C_{5}-C_{4}$~ are arbitrary constants, and $f(r)$ is
a function of ~$r$. The value of the Kretschmann scalar of this
solution is same as that of the previous one $K=4 \Lambda^{2},$
and it has two coordinate singularities at points
\begin{equation}
r_{\pm}=\frac{-D\pm \sqrt{D^{2}+\Lambda C_{3}}}{\Lambda}.
\end{equation}
Using new coordinate ~$\rho=r+\frac{D}{2\Lambda}$,~ the latter
solution \eqref{AdS black hole solution-2} transforms to the $AdS$ black hole solution \eqref{AdS black hole solution}. This also can be achieved by choosing ~$D=0$ and $C_{3}=-b$.

%%%%%%%%%%%%%%%%%%%%%%%%%%%%%%%%%%%%%%%%%%%%
\subsubsection {Black hole solutions}

For negative $\Lambda$, by assuming $N(r)=sinh(\sqrt{-\Lambda}~r-b)$ and $C_{4}=-\Lambda$, the solution \eqref{Solution for non-zero Lambda} reduces to the following black hole solution:
\begin{equation}\label{Black hole solution for negative Lambda}
ds^2=-sinh^{2}(\sqrt{-\Lambda}~r-b)dt^2 +dr^2,
\end{equation}
\begin{equation} \nonumber
\eta_{0}(r)=C_{2}~sinh(\sqrt{-\Lambda}~r-b),~~~~~~~~~~~~~~~~~~~   \eta_{1}(r)=0,~~~~~~~~~~~~~~~~~~~~~~~~~~~~~~~~~~~~~~~~~~~~~~~~~~~~~~~~~~~~
\end{equation}
\begin{equation} \label{L32}
~~~~~~~~\eta_{2}(r)=-\frac{C_{2}}{\sqrt{-\Lambda}}~cosh(\sqrt{-\Lambda}~r-b) +C_{1},~~~~~~~~~~~~~~~~
\eta_{3}(r)=-\frac{C_{2}}{k} \sqrt{-\Lambda} ~cosh(\sqrt{-\Lambda}~r-b),~~~~~~~~~~~~~~~~~~~~~~~~~~~~~~~~~~~~~~~~~~~
\end{equation}
\begin{equation} \nonumber
~~~~~~~~\omega(r)=C_{3} ~dt +f(r) ~dr,~~~~~~~~~~~~~~~~~~~~~~~~~~~~ 
A(r)=\frac{k}{\Lambda} \Big((C_{3}-\sqrt{-\Lambda} ~cosh(\sqrt{-\Lambda}~r-b) ) ~dt +f(r) ~dr \Big),~~~~~~
\end{equation}
where $b$ is an arbitrary constant.

For positive $\Lambda$, by assuming $N(r)=sin(\sqrt{\Lambda}~r-b)$ and $C_{4}=\Lambda$, the solution \eqref{Solution for non-zero Lambda} reduces to the following black hole solution:
\begin{equation}\label{Black hole solution for positive Lambda}
ds^2=-sin^{2}(\sqrt{\Lambda}~r-b)dt^2 +dr^2,
\end{equation}
\begin{equation} \nonumber
~\eta_{0}(r)=C_{2}~sin(\sqrt{\Lambda}~r-b),~~~~~~~~~~~~~~~~~~~~~~   \eta_{1}(r)=0,~~~~~~~~~~~~~~~~~~~~~~~~~~~~~~~~~~~~~~~~~~~~~~~~~~~~~~~~~~~~
\end{equation}
\begin{equation}
~~~~~~~~\eta_{2}(r)=\frac{C_{2}}{\sqrt{\Lambda}}~cos(\sqrt{\Lambda}~r-b) +C_{1},~~~~~~~~~~~~~~~~
\eta_{3}(r)=-\frac{C_{2}}{k} \sqrt{\Lambda} ~cos(\sqrt{\Lambda}~r-b),~~~~~~~~~~~~~~~~~~~~~~~~~~~~~~~~~~~~~~~~~~~
\end{equation}
\begin{equation} \nonumber
~~~\omega(r)=C_{3} ~dt +f(r) ~dr,~~~~~~~~~~~~~~~~~~~~~~~~~~ 
A(r)=\frac{k}{\Lambda} \Big((C_{3}-\sqrt{\Lambda} ~cos(\sqrt{\Lambda}~r-b) ) ~dt +f(r) ~dr \Big).~~~~~~
\end{equation}
Both of the solutions \eqref{Black hole solution for negative Lambda} and \eqref{Black hole solution for positive Lambda} have coordinate singularities at
\begin{equation}
r=\frac{b}{\sqrt{|\Lambda|}},
\end{equation}
which can be removed by suitable coordinate transformations, because of their finite Ricci and Kretschmann scalars.

%%%%%%%%%%%%%%%%%%%%%%%%%%%%%%%%%%%%%%%%%%%%%%%%%%%%%%%%%%
\subsection {Radial solutions for $\Lambda=0$}

As previously discussed, for $\Lambda=0$, the non-semi-simple extension of the Poincar\'{e}
algebra $\mathcal{N}$, reduces to the centrally extended
Poincar\'{e} algebra. Then, the (1+1)-dimensional gauge
symmetric gravity model \eqref{Main action} reduces to the following
central extension of Poincar\'{e} gauge symmetric action
\cite{1Cangemi,Jackiw,2Cangemi}:
\begin{equation} \label{2D gravity for zero Lambda}
I= \int d^{2}x ~  \epsilon^{ij}~ \Big\{ \eta_{a}~
\Big(\partial_{i} e_{j}^{~a} + \epsilon^{ab}~ e_{ib}~ \omega_{j}
\Big) + \eta_{2}~\partial_{i} \omega_{j} +
\eta_{3}~\Big(\partial_{i} A_{j} + \frac{1}{2}k~\epsilon^{ab}~
e_{ia}~ e_{jb}\Big) \Big\}.
\end{equation}
We use the Ansatz \eqref{Metric Ansatz} to solve the equations of motions \eqref{e.o.m.1}-\eqref{e.o.m.2} by inserting $\Lambda=0$ in them, and obtain the following solution:
\begin{equation} \nonumber
M^2(r)= \Big(\frac{1}{D_{1}}\frac{dN(r)}{dr}\Big)^{2},~~~~~~~~~~~~~  
\eta_{0}(r)=-\frac{k D_{2}}{D_{1}}
N(r),~~~~~~~~~~~   \eta_{1}(r)=0,~~~~~~~~~~~~~~~~~~~~
\end{equation}
\begin{equation} \label{Solution for zero Lambda}
\eta_{2}(r)=\frac{k D_{2}}{2 {D_{1}^{2}}} N^{2}(r) +D_{3},~~~~~~~~~~
\eta_{3}(r)=D_{2},~~~~~~~~~~~~~~~~~~~~~~~~~~~~~~~~~~~~~~~~~~~~~~~~~~~~~~
\end{equation}
\begin{equation} \nonumber
\omega(r)=D_{1}  dt,~~~~~~~~~~~~~~~~~~~~~~~~~~ 
A(r)=\Big(\frac{k }{2 {D_{1}}} N^{2}(r) +D_{4} \Big) ~dt +g(r) ~dr,~~~~~~~~~~~~~~~~
\end{equation}
where $D_{1}$, $D_{2}$, $D_{3}$ and $D_{4}$ are arbitrary constants, and $N(r)$, $g(r)$
are arbitrary functions of ~$r$.
The solution \eqref{Solution for zero Lambda} describes a Ricci-flat space-time with zero scalar curvature ($\mathcal{R} =0$).
For $N^2(r)=2 D_{1} r-D_{5}$, the metric in solution \eqref{Solution for zero Lambda} reduces to the following Schwarzschild-type metric:
\begin{equation}\label{Schwarzschild-type metric-2}
ds^{2}=-(2 D_{1} r-D_{5}) dt^{2}+\frac{dr^{2}}{2 D_{1} r-D_{5}},~~~~~~~
\end{equation}
where $D_{5}$ is an arbitrary constant.
The metric \eqref{Schwarzschild-type metric-2} has a
coordinate singularity at
\begin{equation}
r= \frac{D_{5}}{2 D_{1}}.
\end{equation}
For $N^2(r)= (D_{1} r-D_{6})^{2}$, the metric in solution \eqref{Solution for zero Lambda} turns out to be as follows:
\begin{equation}\label{First exact metric}
ds^{2}=-(D_{1} r-D_{6})^{2} dt^{2} +dr^{2},~~~~~~~
\end{equation}
and has a coordinate singularity at
\begin{equation}
r= \frac{D_{6}}{D_{1}},
\end{equation}
where $D_{6}$ is an arbitrary constant.
In section 4, we study the gauged Wess-Zumino-Witten model, and show that the solution \eqref{Solution for zero Lambda} is an exact solution to the string theory, and specially the latter metric solution \eqref{First exact metric} describes an exact (1+1)-dimensional Ricci-flat black hole in string theory.

%%%%%%%%%%%%%%%%%%%%%%%%%%%%%%%%%%%%%%%%%%%%%%%%%%%%%%%%%%
\subsection  {Friedmann-Robertson-Walker (FRW) solutions}

Cosmology of the two-dimensional Jackiw-Teitelboim gravity model is studied in Ref. \cite{Cadoni}. Moreover, some cosmological solutions of the string-inspired gravity coupled to the matter field, both for dust-filled and radiation-filled space-times have been discussed in Ref. \cite{Mann}.
Here, to obtain some cosmological solutions for the equations of motions \eqref{e.o.m.1}-\eqref{e.o.m.2}, we use Friedmann-Robertson-Walker metric as follows:
\begin{equation}\label{FRW metric}
ds^2=-dt^2+ a^2(t) \frac{dr^2}{1- \kappa r^2},
\end{equation}
where $a(t)$ is the scale factor and describes the expansion of
the world, and $\kappa$ is a constant which can be equal to $-1,
0$ or $+1$ only. In 1+1 dimensions, one can use the following coordinate transformation:
\begin{equation}
r \rightarrow \frac{1}{\sqrt{-\kappa}}~sinh(\sqrt{-\kappa}~r)    ~~~~~~~~ for  ~~~~~~~  \kappa<0
\end{equation}
\begin{equation}
r \rightarrow \frac{1}{\sqrt{\kappa}}~sin(\sqrt{\kappa}~r)    ~~~~~~~~~~~~~~ for  ~~~~~~~~  \kappa>0
\end{equation}
to write the metric \eqref{FRW metric} as follows:
\begin{equation}\label{FRW metric in 2D}
ds^2=-dt^2 +a^2(t) dr^2.
\end{equation}
Using the Ansatz \eqref{FRW metric in 2D} for the metric in the equations \eqref{e.o.m.1}-\eqref{e.o.m.2} and after some calculations, one can obtain three different solutions corresponding to the negative, positive or zero values of $\Lambda$, as follows:

%%%%%%%%%%%%%%%%%%%%%%%%%%%%%%%%%%%%%%%%%%
\subsubsection{FRW Solution for $\Lambda<0$}

We have the following solution for the negative $\Lambda$:
\begin{equation} \nonumber
a(t)=\frac{\dot{a}(0)}{\sqrt{-\Lambda}}sin(\sqrt{-\Lambda}~t)
+a(0)cos(\sqrt{-\Lambda}~t), 
~~~~~~~~~~~~~~ 
\end{equation}
\begin{equation} \nonumber
\omega(t,r)=\frac{\Lambda}{k} \Big(h(t,r)dt +s(t,r) dr \Big),~~~~~~~~~ 
A(t,r)=h(t,r)dt + \Big( s(t,r)+\frac{k}{\sqrt{-\Lambda}} \xi_{1}(t) \Big) dr,
\end{equation}
\begin{equation} \label{FRW solution for negative Lambda}
\eta_{0}(r)=\frac{d g(r)}{d r},~~~~~~~~~~~~~~~~~~~~~~~~~~~~~~~~
\eta_{1}(t,r)= -\sqrt{-\Lambda}~\xi_{1}(t) g(r),~~~~~~~~~~~~~~~~~~~~~~
\end{equation}
\begin{equation}   \nonumber
\eta_{2}(t,r)=a(t) g(r),~~~~~~~~~~~~~~~~~~~~~~~~~~~
\eta_{3}(t,r)=-\frac{\Lambda}{k} a(t) g(r),~~~~~~~~~~~~~~~~~~~~~~~~~~~~~~~~~
\end{equation}
where
\begin{equation} \nonumber
\xi_{1}(t)=a(0)~sin(\sqrt{-\Lambda}~t)-\frac{\dot{a}(0)}{\sqrt{-\Lambda}}~cos(\sqrt{-\Lambda}~t),~~~~~~~~~~
s(t,r)=\int dt\frac{\partial h(t,r)}{\partial r},~~~~~~~~
\end{equation}
\begin{equation}
g(r)=C_{1}~cosh(\sqrt{\hat{\lambda}}~r)-C_{2}~sinh(\sqrt{\hat{\lambda}}~r), ~~~~~~~~~~~~~~~~~~~~~~    
\hat{\lambda}=\dot{a}^{2}(0)-\Lambda~a^{2}(0),~~~~
\end{equation}
$C_{1}$ and $C_{2}$ are constants, $h(t,r)$ is an arbitrary function of
~$t$ and $r$, and dot denotes derivative with respect to the timelike
coordinate $t$. 
$a(0)$ and $\dot{a}(0)$ in solution \eqref{FRW solution for negative Lambda} are the initial values of scale factor $a(t)$ and its time derivative $\dot{a}(t)$ at $t=0$, respectively.
Because the fields in the solution \eqref{FRW solution for negative Lambda} are functions
of the radial coordinate $r$, this solution is not a homogenous
solution. In order to obtain a homogeneous solution, $h(t,r)$ must be $r$-independent
$h(t,r)=h_{0}(t)$ and $C_{1}=C_{2}=0$, where $h_{0}(t)$ is a function of timelike coordinate $t$ only. Then, by this choice,
all of the fields will be functions of the coordinate
$t$ only, and spatial homogeneity will be achieved. 
This solution will collapse at
\begin{equation}
t=\frac{1}{\sqrt{-\Lambda}}~arctan\Big(-\sqrt{-\Lambda}~ \frac{a(0)}{\dot{a}(0)}~\Big).
\end{equation}
The Hubble parameter $H(t)$ for this solution is as follows:
\begin{equation}
H(t) \equiv \frac{\dot{a}(t)}{a(t)} = \sqrt{-\Lambda} \Big(
\frac{\dot{a}(0)~cos(\sqrt{-\Lambda}~t)
-\sqrt{-\Lambda}~a(0)~sin(\sqrt{-\Lambda}~t)}{\dot{a}(0)~sin(\sqrt{-\Lambda}~t)
+\sqrt{-\Lambda}~a(0)~cos(\sqrt{-\Lambda}~t)}\Big).
\end{equation}
Using $\ddot{a}(t)=\Lambda a(t)$, the deceleration parameter
$q(t)$ can be obtained as follows:
\begin{equation}
q(t) \equiv -\frac{a(t) \ddot{a}(t)}{\dot{a}^{2}(t)} 
= \Big( \frac{\dot{a}(0)~sin(\sqrt{-\Lambda}~t)
+\sqrt{-\Lambda}~a(0)~cos(\sqrt{-\Lambda}~t)}{\dot{a}(0)~cos(\sqrt{-\Lambda}~t)
-\sqrt{-\Lambda}~a(0)~sin(\sqrt{-\Lambda}~t)} \Big)^{2},
\end{equation}
which is obviously positive, and shows that the expansion of the universe is decelerating.

%%%%%%%%%%%%%%%%%%%%%%%%%%%%%%%%%%%%%%%%%%
\subsubsection{FRW Solution for $\Lambda>0$}

For positive $\Lambda$, one obtains the following solution:
\begin{equation} \nonumber
a(t)=\frac{\dot{a}(0)}{\sqrt{\Lambda}}sinh(\sqrt{\Lambda}~t)
+a(0)cosh(\sqrt{\Lambda}~t), 
~~~~~~~~~~~~~~ 
\end{equation}
\begin{equation} \nonumber
\omega(t,r)=\frac{\Lambda}{k} \Big(\hat{h}(t,r)dt +\hat{s}(t,r) dr \Big),~~~~~~~~~ 
A(t,r)=\hat{h}(t,r)dt + \Big( \hat{s}(t,r)+\frac{k}{\sqrt{\Lambda}} \xi_{2}(t) \Big) dr,
\end{equation}
\begin{equation} \label{FRW solution for positive Lambda}
\eta_{0}(r)=\frac{d \hat{g}(r)}{d r},~~~~~~~~~~~~~~~~~~~~~~~~~~~~~~~~
\eta_{1}(t,r)= \sqrt{\Lambda}~\xi_{2}(t) \hat{g}(r),
~~~~~~~~~~~~~~~~~~~~~~~~
\end{equation}
\begin{equation}   \nonumber
\eta_{2}(t,r)=a(t) \hat{g}(r),~~~~~~~~~~~~~~~~~~~~~~~~~~~
\eta_{3}(t,r)=-\frac{\Lambda}{k} a(t) \hat{g}(r),~~~~~~~~~~~~~~~~~~~~~~~~~~~~~~
\end{equation}
where
\begin{equation} \nonumber
\hat{s}(t,r)=\int dt\frac{\partial \hat{h}(t,r)}{\partial r},~~~~~~~~~~~~~~~~~
\xi_{2}(t)=a(0)~sinh(\sqrt{\Lambda}~t)+\frac{\dot{a}(0)}{\sqrt{\Lambda}}~cosh(\sqrt{\Lambda}~t),
\end{equation}
\begin{equation}
\hat{g}(r)=  \left\{ \begin{tabular}{cc}
$D_{1} r + D_{2}$ ~~~~~~~~~~~~~~~~~~~~~~~~~~~~~~~~~~~~~~~~~~~~    $\hat{\lambda}=0$    \\
$D_{1}~cosh(\sqrt{\hat{\lambda}}~r)+D_{2}~sinh(\sqrt{\hat{\lambda}}~r)$ ~~~~~~~~~~~~~  $\hat{\lambda}>0$        \\
$D_{1}~cos(\sqrt{-\hat{\lambda}}~r)+D_{2}~sin(\sqrt{-\hat{\lambda}}~r)$ ~~~~~~~~~~~~~  $\hat{\lambda}<0$        \\
\end{tabular} \right\},~~~~~
\hat{\lambda}=\dot{a}^{2}(0)-\Lambda~a^{2}(0),
\end{equation}
$D_{1}$ and $D_{2}$ are constants, and $\hat{h}(t,r)$ is an arbitrary function. This solution is not homogenous, and as the previous solution, in order to have a homogenous
solution we must put $\hat{h}(t,r)=\hat{h}_{0}(t)$ and $D_{1}=D_{2}=0$. 
This solution will collapse for $|\frac{\dot{a}(0)}{a(0)}| \geqslant \sqrt{\Lambda}$, at
\begin{equation}
t=\frac{1}{\sqrt{\Lambda}}~arctanh\Big(-\sqrt{\Lambda}~ \frac{a(0)}{\dot{a}(0)}~\Big),
\end{equation}
but for $|\frac{\dot{a}(0)}{a(0)}| < \sqrt{\Lambda}$, it does not collapse.
The Hubble parameter $H(t)$ for this solution is as follows:
\begin{equation}
H(t) \equiv \frac{\dot{a}(t)}{a(t)} = \sqrt{\Lambda} \Big(
\frac{\dot{a}(0)~cosh(\sqrt{\Lambda}~t)
	+\sqrt{\Lambda}~a(0)~sinh(\sqrt{\Lambda}~t)}{\dot{a}(0)~sinh(\sqrt{\Lambda}~t)
	+\sqrt{\Lambda}~a(0)~cosh(\sqrt{\Lambda}~t)}\Big).
\end{equation}
Using $\ddot{a}(t)=\Lambda a(t)$, the deceleration parameter
$q(t)$ can be obtained as follows:
\begin{equation}
q(t) \equiv -\frac{a(t) \ddot{a}(t)}{\dot{a}^{2}(t)}
=-\Big( \frac{\dot{a}(0)~sinh(\sqrt{\Lambda}~t)
	+\sqrt{\Lambda}~a(0)~cosh(\sqrt{\Lambda}~t)}{\dot{a}(0)~cosh(\sqrt{\Lambda}~t)
	+\sqrt{\Lambda}~a(0)~sinh(\sqrt{\Lambda}~t)} \Big)^{2},
\end{equation}
which is obviously negative. Note that for $\dot{a}(0)=\sqrt{\Lambda}~a(0)$, the scale factor of the solution \eqref{FRW solution for positive Lambda} has the following exponential form:
\begin{equation}    
a(t)=a(0)e^{\sqrt{\Lambda}~t},
\end{equation}
and leads to a constant Hubble parameter $H=\sqrt{\Lambda}$ and negative deceleration parameter $q=-1$, which means that the expansion of the universe is accelerating.

%%%%%%%%%%%%%%%%%%%%%%%%%%%%%%%%%%%%%%%%
\subsubsection{FRW Solution for $\Lambda=0$}

For $\Lambda=0$, we obtain the following solution:
\begin{equation} \nonumber
a(t)=\dot{a}(0) t +a(0), ~~~~~~~~~~~~~~~~~~~~~~~~~~~~~~~~~~~~~~~~~~~~~~~~~~~~~~~~~~~
\end{equation}
\begin{equation} \nonumber
\omega(t,r)=-\dot{a}(0) dr,~~~~~~~~ 
A(t,r)=\bar{h}(t,r)dt + \Big\{ \bar{s}(t,r)+k \Big(\frac{1}{2}\dot{a}(0) t^{2} +a(0) t +E_{3}\Big) \Big\} dr,
\end{equation}
\begin{equation} \label{FRW solution for zero Lambda}
\eta_{0}(r)=-\frac{d \bar{g}(r)}{d r},~~~~~~
\eta_{1}(t,r)=-\dot{a}(0) \bar{g}(r), ~~~~~~~~~~~~~~~~~~~~~~~~~~~~~~~~~~~~~~~~~~~~~~~~~~
\end{equation}
\begin{equation}   \nonumber
\eta_{2}(t,r)=-a(t) \bar{g}(r),~~~~~~
\eta_{3}(t,r)=0, ~~~~~~~~~~~~~~~~~~~~~~~~~~~~~~~~~~~~~~~~~~~~~~~~~~~~~~~~~~~~~~~~
\end{equation}
where
\begin{equation} \nonumber
\bar{s}(t,r)=\int dt\frac{\partial \bar{h}(t,r)}{\partial r},~~~~~~~~
\bar{g}(r)= E_{1}~cosh\Big(\dot{a}(0)~r\Big)+E_{2}~sinh\Big(\dot{a}(0)~r\Big),~~~~~~~~~~~~
\end{equation}
$E_{1}$, $E_{2}$ and $E_{3}$ are arbitrary constants, and  $\bar{h}(t,r)$ is an arbitrary function of $t$ and $r$. To obtain a homogenous
solution, $\bar{h}(t,r)$ must be independent of coordinate $r$ \Big($\bar{h}(t,r)=\bar{h}_{0}(t)$\Big), and also we must have $E_{1}=E_{2}=0$. 
This solution obviously will collapse at
\begin{equation}
t= -\frac{a(0)}{\dot{a}(0)}.
\end{equation}
The Hubble parameter $H(t)$ and the deceleration parameter $q(t)$ for this
solution can be obtained as follows:
\begin{equation}
H(t)= \frac{\dot{a}(0)}{\dot{a}(0) t +a(0)},~~~~~~~~~~~~~~ q(t)=0,
\end{equation}
which show that expansion of the universe is without acceleration.
In the next section, by studying the gauged Wess-Zumino-Witten model, we show that the cosmological metric solution \eqref{FRW solution for zero Lambda} is also an exact solution to the string theory.

%%%%%%%%%%%%%%%%%%%%%%%%%%%%%%%%%%%%%%%%%%%%%%%%%%%%%%%%%%
\section  {$\frac{\mathbf{A}_{\mathbf{4,8}}}{\mathbf{A}_{\mathbf{1}}\otimes \mathbf{A}_{\mathbf{1}}}$ gauged Wess-Zumino-Witten model}

As we have explained in the introduction of this paper, it has been shown that two-dimensional string-inspired gravity model \cite{Witten,Verlinde,1Callan} is equivalent to the gauge theory \eqref{2D gravity for zero Lambda} of the centrally extended Poincar\'{e} algebra (the Maxwell algebra in 1+1 space-time dimensions) \cite{1Cangemi,Jackiw,2Cangemi}. So we anticipate that the gravity model \eqref{2D gravity for zero Lambda} which has the Maxwell symmetry, and the string theory obtained by a gauged WZW model on the Maxwell algebra ($\cong \mathcal{A}_{4,8}$), both have some common properties such as a common solution to them.
In this section, we try to find some exact solutions to the string theory (obtained by gauged Wess-Zumino-Witten model using $\mathcal{A}_{4,8}$) which coincide with the solutions of our (1+1)-dimensional gravity model.
We have mentioned in the footnote 3 that the Maxwell algebra in 1+1 dimensions is isomorphic  to the Lie algebra $\mathcal{A}_{4,8}$ \cite{Patera}:
\begin{equation}\label{Lie algebra A_{4,8}}
[X_{2},X_{3}] = X_{1},~~~~~~ [X_{2},X_{4}] = X_{2},~~~~~~
[X_{3},X_{4}]=-X_{3},
\end{equation}
where $\{X_{i}\}, i=1...4$ are the bases of the Lie algebra.
We display the corresponding Lie group with $\mathbf{A}_{\mathbf{4,8}}$.
Now, we study $G/H$ gauged Wess-Zumino-Witten model to obtain an exact solution to the string theory.
Let the Lie group $G$ be $G=\mathbf{A}_{\mathbf{4,8}}$,
and $H=\mathbf{{A^{\!(I)}_{1}}}\otimes \mathbf{A^{\!(II)}_{1}}$ is its subgroup which is a direct product of two one-dimensional Abelian non-compact Lie groups $\mathbf{{A^{\!(I)}_{1}}}$ and $\mathbf{{A^{\!(II)}_{1}}}$ generated by the bases $X_{1}$ and $X_{4}$, respectively. If $g$ is an element of the Lie group $G$, then the Wess-Zumino-Witten
action can be written as follows \cite{2Witten}:
\begin{equation}\label{Wess-Zumino-Witten action}
L(g)=\frac{k}{4\pi} \int_{\Sigma} d^{2}\!z~ \langle g^{-1} \partial g,
g^{-1} \bar{\partial} g \rangle -\frac{k}{24\pi} \int_{B}d^{3}\!\sigma \epsilon^{\alpha\beta\gamma} \langle g^{-1} \partial_{\alpha} g, [g^{-1} \partial_{\beta} g, g^{-1} \partial_{\gamma} g] \rangle,
\end{equation}
where $B$ is a three-dimensional manifold with the coordinates $\sigma^{\mu}=\{z,\bar{z},y\}$, and $\Sigma$ is its boundary with the local complex coordinates $\{z,\bar{z}\}$. Furthermore, we use the notations $\partial=\frac{\partial}{\partial z}$, $\bar{\partial}=\frac{\partial}{\partial \bar{z}}$ and $\partial_{\mu}=\frac{\partial}{\partial \sigma^{\mu}}$, and $d^{2}\!z$ and $d^{3}\!\sigma$ denote the measures $\vert dz d\bar{z} \vert$ and $d^{2}\!z dy$, respectively. By introducing the gauge fields $\mathbf{A},
\mathbf{\bar{A}}$ which takes their values in the Lie algebra
of $H$, the gauged Wess-Zumino-Witten action having the local
axial symmetry $g\longrightarrow hgh,~\mathbf{A}\longrightarrow h(\mathbf{A}+\partial)h^{-1}$ and $\mathbf{\bar{A}}\longrightarrow h^{-1}(\mathbf{\bar{A}}-\bar{\partial})h~(h\in H)$ can be written as follows \cite{Gawedzki}:
\begin{equation}\label{gauged Wess-Zumino-Witten action}
L(g,\mathbf{A})=L(g) +\frac{k}{2\pi} \int_{\Sigma} d^{2}z~
\Big(\langle \mathbf{A}, \bar{\partial} g~g^{-1} \rangle +
\langle \mathbf{\bar{A}}, g^{-1} \partial g \rangle +\langle \mathbf{A}, \mathbf{\bar{A}} \rangle + \langle g^{-1} \mathbf{A} g, \mathbf{\bar{A}} \rangle \Big).
\end{equation}
Here, we consider the Lie algebra $\mathcal{H}=\mathcal{A}^{\!(I)}_{1} \oplus \mathcal{A}^{\!(I\!I)}_{1}$ valued gauge fields $\mathbf{A}$ and $\mathbf{\bar{A}}$ as follows:
\begin{equation}\label{gauge fields}
\mathbf{A}=A_{1} X_{1}+A_{2} X_{4},~~~~~~~~~~~
\mathbf{\bar{A}}=\bar{A}_{1} X_{1} +\bar{A}_{2} X_{4}.
\end{equation}
We parameterize $G=\mathbf{A}_{\mathbf{4,8}}$ group by the following group
element:
\begin{equation}\label{group element}
g=e^{aX_{1}}~ e^{bX_{2}}~ e^{uX_{3}}~ e^{vX_{4}},
\end{equation}
where $a,b,u,v\in \mathbb{R}$.
Then, using the group element \eqref{group element}, and the following non-degenerate ad-invariant bilinear quadratic form \cite{Nappi} on the Lie algebra $\mathcal{A}_{4,8}$ \eqref{Lie algebra A_{4,8}}:
\begin{equation}\label{bilinear quadratic form}
\langle X_{1},X_{4}\rangle =\alpha,~~~~~
\langle X_{2},X_{3}\rangle =-\alpha,~~~~~
\langle X_{4},X_{4}\rangle =\beta,~~~~~
\end{equation}
one can rewrite the Wess-Zumino-Witten
action \eqref{Wess-Zumino-Witten action} as follows:
\begin{equation}
L(g)=\frac{k}{4\pi} \!\!\int_{\Sigma}\!\!\! d^{2}\!z \Big( \alpha(\partial a \bar{\partial} v +\bar{\partial}a \partial v) +\alpha u(\partial b \bar{\partial} v +\bar{\partial} b\partial v ) -\alpha (\partial b \bar{\partial} u +\bar{\partial} b\partial u ) +\beta \partial v \bar{\partial}v \Big)
-\frac{k\alpha}{4\pi} \!\!\int_{\Sigma}\!\!\! d^{2}\!z~ v(\partial b \bar{\partial} u -\bar{\partial} b\partial u ),
\end{equation}
where $\alpha$ and $\beta$ are arbitrary constants ($\alpha$ should be nonzero in order that  the ad-invariant bilinear quadratic form \eqref{bilinear quadratic form} be non-degenerate).
We gauge $\mathbf{{A^{\!(I)}_{1}}} \otimes \mathbf{{A^{\!(II)}_{1}}}$ subgroup generated infinitesimally by\footnote{The infinitesimal gauge transformation $\delta g=\epsilon_{1}(X_{1} g+g X_{1}) +\epsilon_{2}(X_{4} g+g X_{4})$ can be obtained by substituting $h=e^{\epsilon_{1}X_{1}+\epsilon_{2}X_{4}}$ in the axial symmetry $g\longrightarrow hgh$.} $\delta g=\epsilon_{1}(X_{1} g+g X_{1}) +\epsilon_{2}(X_{4} g+g X_{4})$ which using \eqref{group element} gives the following transformations of the group parameters:
\begin{equation}
\delta a=2\epsilon_{1},~~~~~ \delta b=-b\epsilon_{2},~~~~~ \delta u=u\epsilon_{2},~~~~~ \delta v=2\epsilon_{2},
\end{equation}
together with the following transformations of the gauge fields:
\begin{equation}
\delta A_{i}=-\partial\epsilon_{i},~~~~~ \delta \bar{A}_{i}=-\bar{\partial}\epsilon_{i},~~~~~ (i=1,2)
\end{equation}
where $\epsilon_{1}$ and $\epsilon_{2}$ are gauge parameters.
Then, the resultant gauged WZW action is as follows:
\begin{equation}\nonumber
L(g,\mathbf{A})=L(g) +\frac{k}{2\pi} \!\!\int_{\Sigma}\!\!\! d^{2}z~
\Big\{\alpha \bar{\partial} v A_{1} +\Big(\alpha(\bar{\partial} a+b\bar{\partial} u -bu\bar{\partial}v)+\beta\bar{\partial}v \Big) A_{2}  +\alpha \partial v \bar{A}_{1} + \Big(\alpha(\partial a+u\partial b)+\beta \partial v\Big) \bar{A}_{2} 
\end{equation}
\begin{equation}\label{gauged WZW model}
+2\alpha A_{1} \bar{A}_{2} +2\alpha A_{2} \bar{A}_{1}
+(2\beta-\alpha bu) A_{2} \bar{A}_{2} \Big\},
\end{equation}
Variations of the action \eqref{gauged WZW model} with respect to the  gauge fields $A_{i}$ and $\bar{A}_{i}$ ($i=1,2$) gives the following equations:
\begin{equation}\nonumber
A_{1}=-\frac{1}{2}(\partial a+u\partial b+\frac{1}{2}bu \partial v),~~~~~~~~
A_{2}=-\frac{1}{2}\partial v,
\end{equation}
\begin{equation}
\bar{A}_{1}=-\frac{1}{2}(\bar{\partial} a+b\bar{\partial} u-\frac{1}{2}bu \bar{\partial} v),~~~~~~~~
\bar{A}_{2}=-\frac{1}{2}\bar{\partial} v,
\end{equation}
using which one eliminates the gauge fields in \eqref{gauged WZW model}, and obtains the following gauged Wess-Zumino-Witten action:
\begin{equation}\label{GWZW-1}
L(g,\mathbf{A})=\frac{k}{4\pi} \!\!\int_{\Sigma}\!\!\! d^{2}\!z \Big( \frac{1}{2}\alpha bu\partial v \bar{\partial} v +\alpha\partial v (u \bar{\partial} b -b\bar{\partial} u ) -\alpha (\partial b \bar{\partial} u +\bar{\partial} b\partial u ) \Big)
-\frac{k\alpha}{4\pi} \!\!\int_{\Sigma}\!\!\! d^{2}\!z~ v(\partial b \bar{\partial} u -\bar{\partial} b\partial u ).
\end{equation}
This action is independent of the parameter $a$, and now we can fix the gauge by setting $b=u$, such that the gauged WZW action \eqref{GWZW-1} turns out to be
\begin{equation}\label{GWZW-2}
L(g,\mathbf{A})=-\frac{k\alpha}{2\pi}\int_{\Sigma} d^{2}z~
\Big( -\frac{1}{4} u^{2} \partial v\bar{\partial}v +\partial u \bar{\partial}u \Big).
\end{equation}
Using the following field redefinition:
\begin{equation}
u(r)=\frac{\hat{N}(r)}{\hat{D}} ,~~~~~~~~
v(t)=2\hat{D} t,
\end{equation}
the gauged WZW action \eqref{GWZW-2} becomes
\begin{equation}\label{GWZW-3}
L(g,\mathbf{A})=-\frac{k\alpha}{2\pi}\int_{\Sigma} d^{2}z~
\Big( -\hat{N}^{2}(r) \partial t \bar{\partial}t +\Big(\frac{1}{\hat{D}}\frac{d\hat{N}(r)}{dr}\Big)^{2}\partial r \bar{\partial}r \Big),
\end{equation}
where $\hat{D}$ is a constant and $\hat{N}(r)$ is an arbitrary function of the spatial coordinate $r$.
Then, the obtained gauged WZW action \eqref{GWZW-3} describes a string
propagating in a space-time with the following metric
\begin{equation}\label{Exact black hole in string theory}
ds^{2}=-\hat{N}^{2}(r) dt^{2} +\Big(\frac{1}{\hat{D}}\frac{d\hat{N}(r)}{dr}\Big)^{2} dr^{2}.
\end{equation}
By assuming $\hat{D}=D_{1}$ and $\hat{N}(r)=N(r)$, the metric solution \eqref{Exact black hole in string theory} is precisely same as the metric in \eqref{Solution for zero Lambda} obtained as a solution of our gravity model \eqref{Main action}. As we have discussed before, by assuming $\hat{N}^{2}(r)=(D_{1}r-D_{6})^{2}$ the metric \eqref{Exact black hole in string theory} reduces to the metric \eqref{First exact metric} which describes an exact Ricci-flat black hole in string theory.
Now, for obtaining the dilaton field, we consider the following one-loop beta function equations in 1+1 dimensions \cite{2Callan}:
\begin{equation} \label{beta function equations-1}
R_{\mu\nu}+2\nabla_{\mu}\nabla_{\nu}\phi=0,
\end{equation}
\begin{equation} \label{beta function equations-2}
R+\frac{8}{k^{'}}+4\nabla^{2}\phi-4(\nabla\phi)^{2}=0,~~~~~~~~~~~~~
\end{equation}
where $R$ and $R_{\mu\nu}$ are the scalar curvature and Ricci tensor of the target space, $\phi$ is the dilaton field, and $\frac{8}{k^{'}}$ is the cosmological constant term.
By requiring that the metric \eqref{Exact black hole in string theory} must obey the one-loop beta function equations \eqref{beta function equations-1} and \eqref{beta function equations-2}, we obtain the following relation for the dilaton field using \eqref{beta function equations-1}:
\begin{equation}
\phi(t,r)= \hat{N}(r) (c_{1}~cosh(\hat{D} t)+c_{2}~sinh(\hat{D} t)),
\end{equation}
where $c_{1}$ and $c_{2}$ are some real constants. By substituting $\phi(t,r)$ in \eqref{beta function equations-2}, one can obtain the following relation between the constants $c_{1}$ and $c_{2}$:
\begin{equation}
\hat{D}^{2}({c_{1}}^{2} - {c_{2}}^{2}) = \frac{2}{k^{'}}.
\end{equation}
In the same way, using another field redefinition as follows:
\begin{equation}
u(t)=\frac{\alpha t+\beta}{\alpha},~~~~~~~ v(r)=\alpha r,
\end{equation}
where $\alpha$ and $\beta$ are arbitrary real constants, the gauged WZW action \eqref{GWZW-2} turns out to have the following form:
\begin{equation}
L(g,\mathbf{A})=\frac{k\alpha}{2\pi}\int_{\Sigma} d^{2}z~
\Big(-\partial t \bar{\partial}t + (\alpha t+\beta)^{2} \partial r \bar{\partial}r  \Big),
\end{equation}
which describes a string propagating in a space-time with the following cosmological metric:
\begin{equation}\label{Second exact metric}
ds^{2}=-dt^{2} + (\alpha t+\beta)^{2} dr^{2}.
\end{equation}
By assuming $\alpha=\dot{a}(0)$ and $\beta=a(0)$, this metric is precisely same as the FRW metric \eqref{FRW solution for zero Lambda} which is obtained by solving the equations of motion for our gravity model discussed in the previous section.
In the same way for obtaining the previous dilaton field, we use \eqref{beta function equations-1} and \eqref{beta function equations-2} to obtain the following dilaton field corresponding to the metric \eqref{Second exact metric}:
\begin{equation}
\phi(t,r)=(\alpha t+\beta) (d_{1}~cosh(\alpha r)+d_{2}~sinh(\alpha r)),~~~~~~~~~~~~~  \alpha^{2}({d_{2}}^{2} - {d_{1}}^{2}) = \frac{2}{k^{'}},
\end{equation}
where $d_{1}$ and $d_{2}$ are real constants.
Note that the black hole metric \eqref{First exact metric} converts to the FRW metric \eqref{Second exact metric}, and vice versa, using the following coordinate transformation:
\begin{equation} 
t \rightarrow \hat{r},~~~~~~~~~ r \rightarrow \hat{t}.
\end{equation}

%%%%%%%%%%%%%%%%%%%%%%%%%%%%%%%%%%%%%%%%%%%%%%%%%%%%%%%%%%%%%%%%%%%%%%%%%%%%%%%%%%%%%%%%%%%%%%
\section {\large {\bf Conclusions}}
We have presented a four-dimensional extension of the Poincar\'{e}
algebra in (1+1)-dimensional space-time. Using this algebra, we
have constructed a gauge theory of gravity, which is dual
(canonically transformed) to the $AdS$ gauge theory of gravity,
under special conditions. We have also obtained black
hole and Friedmann-Robertson-Walker (FRW) cosmological solutions of this model. Then, using $\frac{\mathbf{A}_{\mathbf{4,8}}}{\mathbf{A}_{\mathbf{1}}\otimes \mathbf{A}_{\mathbf{1}}}$ gauged Wess-Zumino-Witten action,
we have shown that some of the black hole and cosmological solutions of our gravity model are exact (1+1)-dimensional solutions of string theory. 
In this paper, we have shown that the Ricci-flat ($\mathcal{R}=0$) solutions of our gravity model are also exact solutions to the string theory, only. But, it remains an interesting question to be investigated that if our other solutions of the gravity model (with $\mathcal{R}=2\Lambda$) are also exact solutions to the string theory? Analysis of the constraints of the gravity model \eqref{Main action} and its quantization are some of the interesting open problems which may lead to some desired results. Another interesting problem may be the possibility of obtaining the (1+1)-dimensional gravity model \eqref{Main action} by an appropriate dimensional reduction from a gauge invariant (2+1)-dimensional gravity model (see \cite{Achucarro,Grignani}).

%%%%%%%%%%%%%%%%%%%%%%%%%%%%%%%%%%%%%%%%%%%%%%%%%%%%%%%%%%%%%%%%%%%%%%%%%%%%%%%%%%%%%%%%%%%%%%%%%%%%%%%%%%%%%%%%
\textbf{Acknowledgments:} 

We would like to express our heartfelt gratitude to M.M. Sheikh-Jabbari for his useful comments.
This research was supported by a research fund No. 217D4310 from Azarbaijan Shahid Madani
university.

%%%%%%%%%%%%%%%%%%%%%%%%%%%%%%%%%%%%%%%%%%%%%%%%%%%%%%%%%%%%%%%%%%%%%%%%%%%%%%%%%%%%%%%%%%%%%%%%%%%%%%%%%%%%%%%%

\end{document}